\documentclass[twocolumn,aps,prc,showpacs,superscriptaddress,nofootinbib,floatfix]{revtex4}

\usepackage{graphicx}
\usepackage{dcolumn}
\usepackage{bm}
\usepackage[dvipdfm]{hyperref}

\begin{document}

\title{Neutral Pion Production in Au+Au Collisions at $\sqrt{s_{NN}}$ = 200 GeV}

\affiliation{Argonne National Laboratory, Argonne, Illinois 60439, USA}
\affiliation{University of Birmingham, Birmingham, United Kingdom}
\affiliation{Brookhaven National Laboratory, Upton, New York 11973, USA}
\affiliation{University of California, Berkeley, California 94720, USA}
\affiliation{University of California, Davis, California 95616, USA}
\affiliation{University of California, Los Angeles, California 90095, USA}
\affiliation{Universidade Estadual de Campinas, Sao Paulo, Brazil}
\affiliation{University of Illinois at Chicago, Chicago, Illinois 60607, USA}
\affiliation{Creighton University, Omaha, Nebraska 68178, USA}
\affiliation{Czech Technical University in Prague, FNSPE, Prague, 115 19, Czech Republic}
\affiliation{Nuclear Physics Institute AS CR, 250 68 \v{R}e\v{z}/Prague, Czech Republic}
\affiliation{Institute of Physics, Bhubaneswar 751005, India}
\affiliation{Indian Institute of Technology, Mumbai, India}
\affiliation{Indiana University, Bloomington, Indiana 47408, USA}
\affiliation{Institut de Recherches Subatomiques, Strasbourg, France}
\affiliation{University of Jammu, Jammu 180001, India}
\affiliation{Joint Institute for Nuclear Research, Dubna, 141 980, Russia}
\affiliation{Kent State University, Kent, Ohio 44242, USA}
\affiliation{University of Kentucky, Lexington, Kentucky, 40506-0055, USA}
\affiliation{Institute of Modern Physics, Lanzhou, China}
\affiliation{Lawrence Berkeley National Laboratory, Berkeley, California 94720, USA}
\affiliation{Massachusetts Institute of Technology, Cambridge, MA 02139-4307, USA}
\affiliation{Max-Planck-Institut f\"ur Physik, Munich, Germany}
\affiliation{Michigan State University, East Lansing, Michigan 48824, USA}
\affiliation{Moscow Engineering Physics Institute, Moscow Russia}
\affiliation{City College of New York, New York City, New York 10031, USA}
\affiliation{NIKHEF and Utrecht University, Amsterdam, The Netherlands}
\affiliation{Ohio State University, Columbus, Ohio 43210, USA}
\affiliation{Old Dominion University, Norfolk, VA, 23529, USA}
\affiliation{Panjab University, Chandigarh 160014, India}
\affiliation{Pennsylvania State University, University Park, Pennsylvania 16802, USA}
\affiliation{Institute of High Energy Physics, Protvino, Russia}
\affiliation{Purdue University, West Lafayette, Indiana 47907, USA}
\affiliation{Pusan National University, Pusan, Republic of Korea}
\affiliation{University of Rajasthan, Jaipur 302004, India}
\affiliation{Rice University, Houston, Texas 77251, USA}
\affiliation{Universidade de Sao Paulo, Sao Paulo, Brazil}
\affiliation{University of Science \& Technology of China, Hefei 230026, China}
\affiliation{Shandong University, Jinan, Shandong 250100, China}
\affiliation{Shanghai Institute of Applied Physics, Shanghai 201800, China}
\affiliation{SUBATECH, Nantes, France}
\affiliation{Texas A\&M University, College Station, Texas 77843, USA}
\affiliation{University of Texas, Austin, Texas 78712, USA}
\affiliation{Tsinghua University, Beijing 100084, China}
\affiliation{United States Naval Academy, Annapolis, MD 21402, USA}
\affiliation{Valparaiso University, Valparaiso, Indiana 46383, USA}
\affiliation{Variable Energy Cyclotron Centre, Kolkata 700064, India}
\affiliation{Warsaw University of Technology, Warsaw, Poland}
\affiliation{University of Washington, Seattle, Washington 98195, USA}
\affiliation{Wayne State University, Detroit, Michigan 48201, USA}
\affiliation{Institute of Particle Physics, CCNU (HZNU), Wuhan 430079, China}
\affiliation{Yale University, New Haven, Connecticut 06520, USA}
\affiliation{University of Zagreb, Zagreb, HR-10002, Croatia}

\author{B.~I.~Abelev}\affiliation{University of Illinois at Chicago, Chicago, Illinois 60607, USA}
\author{M.~M.~Aggarwal}\affiliation{Panjab University, Chandigarh 160014, India}
\author{Z.~Ahammed}\affiliation{Variable Energy Cyclotron Centre, Kolkata 700064, India}
\author{A.~V.~Alakhverdyants}\affiliation{Joint Institute for Nuclear Research, Dubna, 141 980, Russia}
\author{B.~D.~Anderson}\affiliation{Kent State University, Kent, Ohio 44242, USA}
\author{D.~Arkhipkin}\affiliation{Brookhaven National Laboratory, Upton, New York 11973, USA}
\author{G.~S.~Averichev}\affiliation{Joint Institute for Nuclear Research, Dubna, 141 980, Russia}
\author{J.~Balewski}\affiliation{Massachusetts Institute of Technology, Cambridge, MA 02139-4307, USA}
\author{O.~Barannikova}\affiliation{University of Illinois at Chicago, Chicago, Illinois 60607, USA}
\author{L.~S.~Barnby}\affiliation{University of Birmingham, Birmingham, United Kingdom}
% ...

\collaboration{STAR Collaboration}\noaffiliation

\date{\today}

\begin{abstract}

The results of mid-rapidity ($0 < y < 0.8$) neutral pion spectra
over an extended transverse momentum range ($1 < p_T < 12$ GeV/$c$)
in $\sqrt{s_{NN}}$ = 200 GeV Au+Au collisions, measured by the STAR
experiment, are presented. The neutral pions are reconstructed from
photons measured either by the STAR Barrel Electro-Magnetic
Calorimeter (BEMC) or by the Time Projection Chamber (TPC) via
tracking of conversion electron-positron pairs. Our measurements are
compared to previously published $\pi^{\pm}$ and $\pi^0$ results.
The nuclear modification factors $R_{\mathrm{CP}}$ and
$R_{\mathrm{AA}}$ of $\pi^0$ are also presented as a function of
$p_T$ . In the most central Au+Au collisions, the binary collision
scaled $\pi^0$ yield at high $p_T$ is suppressed by a factor of
about 5 compared to the expectation from the yield of p+p
collisions. Such a large suppression is in agreement with previous
observations for light quark mesons and is consistent with the
scenario that partons suffer considerable energy loss in the dense
medium formed in central nucleus-nucleus collisions at RHIC.

\end{abstract}

\maketitle

\section{Introduction}

The observation of ``jet quenching'' \cite{starwhitepaper,
phenixwhitepaper} in central Au+Au collisions is one of the most
exciting experimental discoveries at the Relativistic Heavy-Ion
Collider (RHIC). Experimental signature of this observation includes
the suppression of inclusive hadron yields at high transverse
momentum ($p_T$) \cite{starppnorm, highptsuppression} and of
associated $p_T>$2 GeV/$c$ particles on the away-side of a high
$p_T$ trigger hadron \cite{noawayside}. These measurements indicate
that RHIC has produced high energy density matter that is opaque to
high $p_T$ quarks and gluons \cite{starwhitepaper}. Theoretical
calculations based on energy loss of high $p_T$ partons through
gluon radiation can explain the suppression of light quark mesons
\cite{GLV}. Measurements of $\pi^0$ at high $p_T$ provide a
fundamental tool for probing the parton energy loss in the medium
created in central nucleus-nucleus collisions at RHIC. On the other
hand, this medium appears to be transparent to direct photons, of
which the nuclear modification factor $R_{\mathrm{AA}}$ is found to
be approximately unity at high $p_T$ \cite{directphoton}.
Measurement of the $\pi^0$ spectrum over an extended $p_T$ range is
a prerequisite to understand the decay photon background of the
direct photon analysis. This measurement also provides an important
cross-check for other pion measurements at RHIC using different
detectors.

In this article we present the first results for the $\pi^0$ spectra
and nuclear modification factors at mid-rapidity, over a broad $p_T$
region ($1 < p_T < 12$ GeV/$c$) in Au+Au collisions at
$\sqrt{s_{NN}}$ = 200 GeV measured by the STAR experiment. Neutral
pions are reconstructed via the di-photon decay channel. Only the
west half of the STAR Barrel Electro-Magnetic Calorimeter (BEMC)
\cite{bemc} was completed and commissioned to take heavy-ion
collision data in 2004. Measurement of a $\pi^0$ spectrum under the
high multiplicity environment in central Au+Au collisions is
challenging due to the large transverse size of the STAR BEMC towers
($0.05\times0.05$ in $\Delta\eta\times\Delta\phi$), resulting in
high occupancy and appreciable background contamination. The BEMC
provides STAR with a trigger capability on high $p_T$ photons based
on large energy deposition in a single BEMC tower or a tower patch.
These triggered BEMC photons can be used to reliably construct
$\pi^0$ mesons in the high $p_T$ region. However, this is not
possible in the low $p_T$ region as the energy resolution of the
BEMC is not sufficiently good. The STAR Time Projection Chamber
(TPC) \cite{tpc} has been used to reconstruct photons that convert
to electron-positron pairs \cite{starpi0130}. Excellent detection
resolution on $\pi^0$ invariant masses has been achieved from TPC
conversion photons. However, the small photon conversion probability
in the STAR detector system restricts the $p_T$ reach. By combining
BEMC photons from high $p_T$ triggers and conversion photons from
the TPC, we have been able to achieve good invariant mass resolution
on $\pi^0$ reconstruction and measure its spectrum over a broad
range of $p_T$.

\section{Data Analysis}

\subsection{Data Set}

The data used in this analysis were taken during the 2004 RHIC run
for Au+Au collisions at the energy $\sqrt{s_{NN}}$ = 200 GeV. The
primary STAR detectors used for this analysis are the TPC and BEMC.
A Barrel Shower Maximum Detector (BSMD) \cite{bemc} at a depth of 5
radiation lengths ($X_0$) inside the BEMC measures transverse shower
shape and position with higher precision than the BEMC tower. Three
Au+Au data sets were used: $11 \times 10^6$ events selected by a
Minimum-Bias trigger (MB), $17 \times 10^6$ events selected by a
central trigger, and $2.4 \times 10^6$ events selected by a High
Tower trigger (HT). The central trigger corresponds to the highest
$12\%$ charged particle multiplicity events as determined by the
coincidence of the Central Trigger Barrel and the Zero Degree
Calorimeters \cite{startrigger}. The HT trigger, which depends on
pseudo-rapidity, requires that at least one BEMC tower has deposited
transverse energy greater than the HT energy threshold of 3-4 GeV.
The HT trigger enhances selection of events containing high $p_T$
photons, and thus helps to extend our measurement to higher $p_T$.
More details about the STAR trigger system and trigger configuration
can be found in Ref. \cite{startrigger}. In this analysis, the
position of the primary vertex is required to be within $\pm$20 cm
of the center of the STAR TPC along the beam line. This requirement
restricts our conversion photon candidates to mid-rapidity, where
the detector geometry is relatively simple and the material is well
studied for reconstructing conversion photons.

\subsection{Photon Identification}

There are two ways to identify photons in STAR: The STAR BEMC and
BSMD measure photons directly from the electromagnetic shower (EMC
photon); or the STAR TPC reconstructs photon conversion to $e^+/e^-$
pairs (TPC photon) in materials such as the beam pipe, the Silicon
Vertex Tracker (SVT), the Silicon Strip Detector (SSD), and TPC
walls and gas. In total these materials are estimated to be
equivalent to about 0.1$X_0$, with a 10\% uncertainty based on
studies of conversion probability correction. The photon conversion
probability will be discussed in Sec. \ref{sec:efficiency} and
\ref{sec:syserror}.

\begin{figure*}
\includegraphics[width=4.8in]{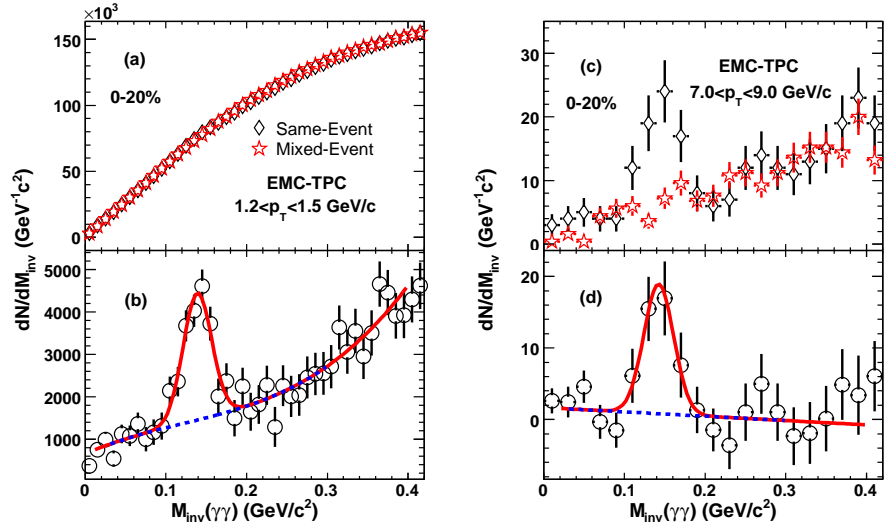}
\caption{\label{fig:emctpc_inv_mass_200gev}The di-photon invariant
mass distributions using the EMC-TPC method in 0-20\% Au+Au
collisions at $\sqrt{s_{NN}}$ = 200 GeV. The solid line is a fit
result of a Gaussian peak plus a background function. The residual
background is shown as a dotted line. Panels (a) and (b) are from a
low $p_T$ bin in MB events; panels (c) and (d) are from a high $p_T$
bin in HT events. Panels (b) and (d) are distributions after
mixed-event background subtraction from panels (a) and (c).}
\end{figure*}

An EMC photon is reconstructed from a single tower. The photon
energy is determined by the tower energy. In MB and central events,
towers with energy greater than 500 MeV and at least 250 MeV higher
than any of their eight surrounding towers are required. The BSMD
hit information is not used due to its expected inefficiency for low
energy photons. The photon position is assumed to be at the center
of the tower. Charged particle contamination is greatly reduced by
projecting TPC tracks into the BEMC and vetoing the first two towers
intersected by the track.

In HT triggered events, BSMD hits are used to separate the two close
decay photons from a single $\pi^0$ decay. The photon positions are
determined from the BSMD hits. If multiple photons are found in the
same tower, the tower energy is split according to the individual
BSMD hit energies. For a photon with energy below the HT threshold,
we require that no TPC track is projected into an area of $\pm$ 0.05
in $\Delta\eta$ and $\pm$ 0.05 in $\Delta\phi$ around the photon
candidate. For a photon above the HT threshold, we require that the
sum of momenta of all charged particle tracks projected to the
surrounding $\Delta\eta-\Delta\phi$ area should be less than 1
GeV/$c$.

\begin{figure}
\includegraphics[width=3in]{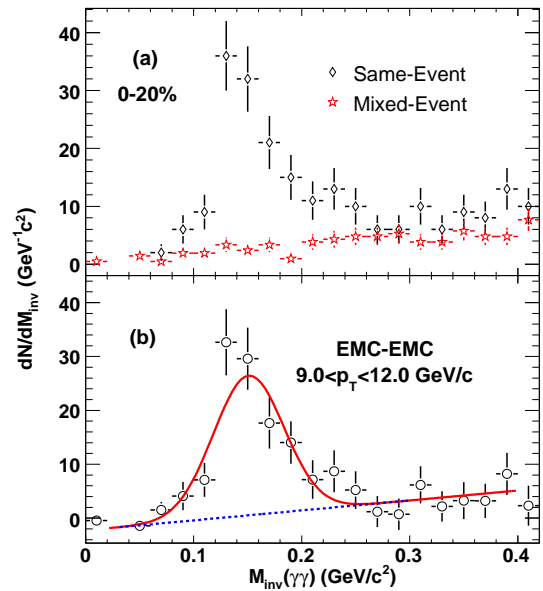}
\caption{\label{fig:emcemc_inv_mass_200gev}The di-photon invariant
mass distributions using the EMC-EMC method in 0-20\% Au+Au
collisions at $\sqrt{s_{NN}}$ = 200 GeV. The solid line is a fit
result of a Gaussian peak plus a background function. The residual
background is shown as a dotted line. Panels (b) is the distribution
after mixed-event background subtraction from panel (a).}
\end{figure}
For TPC photons, we select $e^+/e^-$ candidates via ionization
energy loss $dE/dx$ in the TPC. A number of geometrical cuts are
applied to each $e^+/e^-$ pair to have a topological signature of a
photon conversion. These cuts require that the two tracks originate
from a common secondary vertex within or before entering the TPC
with a small opening angle and a small invariant mass, and that the
reconstructed photon candidate originates from the primary vertex.
The photon momentum is taken as the sum of two daughter track
momenta at the conversion point. This technique has been used in
Au+Au collisions at $\sqrt{s_\mathrm{NN}}$ = 130 GeV, and more
details can be found in Ref. \cite{starpi0130}.

\subsection{$\mathbf{\pi^0}$ Reconstruction}

In MB and central events an EMC photon is paired with a TPC photon
(EMC-TPC), and in HT events pairs of two EMC photons (EMC-EMC) are
also used to reconstruct $\pi^0$'s. Due to the relatively large
tower size of the STAR BEMC, energy determinations for low $p_T$
photons can be contaminated due to the presence of other nearby
photons, neutral particles, and residual charged particles in high
multiplicity Au+Au events. The limited energy resolution of BEMC
towers at low energy (nominal resolution of BEMC towers has been
estimated to be $16\%/\sqrt{E}\oplus1.5\%$) \cite{bemc} also hinders
the accurate measurement of photon energy. As a result, it is
difficult to obtain a clear $\pi^0$ signal at low $p_T$ by
exclusively pairing EMC photons. On the other hand, the relatively
tight geometrical cuts for TPC photon reconstruction select very
clean conversion photon samples. They significantly reduce the
combinatoric background and improve the $\pi^0$ mass resolution. The
EMC-TPC method yields a clear $\pi^0$ signal from 1 GeV/$c$ to
intermediate $p_T$ ($\sim$5 GeV/$c$) in central Au+Au collisions. At
higher $p_T$ above the HT threshold, the EMC photons are less
affected by backgrounds and the EMC-EMC method produces clear
$\pi^0$ signals. Due to its greater efficiency for high $p_T$
photons, the EMC-EMC method is able to extend the measurement to
higher $p_T$.

The mixed-event technique is used to reproduce a combinatoric
background. For the mixed-event distribution, photons from the event
being analyzed are paired with photons from events in an event pool,
in which events are required to have similar multiplicity and
primary vertex position as the one being analyzed. The di-photon
invariant mass distribution after mixed-event background subtraction
is fit to extract the raw $\pi^0$ yield.

Figures \ref{fig:emctpc_inv_mass_200gev} and
\ref{fig:emcemc_inv_mass_200gev} show examples of the di-photon
invariant mass distributions before and after mixed-event background
subtraction for different $\pi^0$ reconstruction methods. In Fig.
\ref{fig:emctpc_inv_mass_200gev} we show the invariant mass
distributions in the $p_T$ regions 1.2-1.5 and 7.0-9.0 GeV/$c$ from
the EMC-TPC method, and Fig. \ref{fig:emcemc_inv_mass_200gev} shows
the invariant mass distributions in the $p_T$ region 9.0-12.0
GeV/$c$ from the EMC-EMC method. All invariant mass distributions
are for the 0-20\% collision centrality bin. These distributions are
fit using a Gaussian plus a polynomial background function. At high
$p_T$, the background is small and can be easily subtracted by
fitting a linear dependence on $M_{inv}$. At low $p_T$, the
signal-to-background ratio is rather small. After mixed-event
background subtraction, a larger residual background is observed and
a 3rd order polynomial function is used to fit the background shape.
Here the normalization factor between same-event and mixed-event is
adjusted for each $p_T$ bin, so that the residual background has a
shape roughly linearly increasing with mass, and can be described by
a polynomial fit. The residual background may come from correlated
photons that are not combinatoric and cannot be reproduced by the
mixed-event technique. Such correlations may arise from
contaminations to EMC photons or from resonance decays to multiple
photons in the final state. The amount of residual background is
strongly centrality and $p_T$ dependent, more pronounced in the most
central events and at lower $p_T$. Figure
\ref{fig:emctpc_inv_mass_200gev}a shows the situation where both
combinatoric and residual backgrounds are most severe; nevertheless
the $\pi^0$ signal can still be observed above the residual
background (Fig. \ref{fig:emctpc_inv_mass_200gev}b).

Several systematic checks have been performed. Firstly, a track
rotation technique was used to generate combinatoric background for
comparison. It rotates the EMC photons by $180^\circ$ in the
azimuthal plane, and mixes them with the photons reconstructed in
the TPC. Secondly, the normalization factor was adjusted and the
invariant mass distribution was re-fit to extract $\pi^0$ yield.
Although these two procedures may significantly change the shape of
residual background, yields extracted using the same function are
consistent with each other. We have also changed the order of
polynomial used for background fitting, as well as the fit range,
and have included the variance in the overall systematic errors.

\begin{figure}
\includegraphics[width=3in]{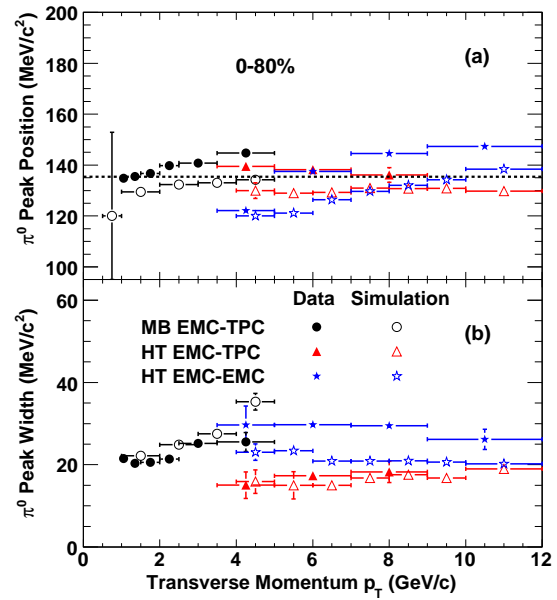}
\caption{\label{fig:peak_pos_width_200gev}The $\pi^0$ invariant mass
peak positions (a) and peak widths (b) as a function of $p_T$ in
0-80\% Au+Au collisions at $\sqrt{s_{NN}}$ = 200 GeV. Corresponding
results from $\pi^0$ embedded simulation are shown for comparison.}
\end{figure}

\begin{figure*}
\includegraphics[width=5in]{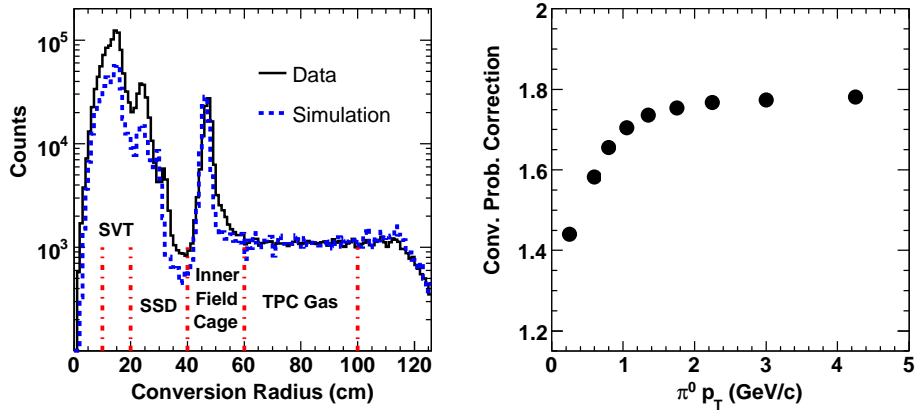}
\caption{\label{fig:geo_effect}Left: Photon conversion point radius
distributions from real data and MC simulation in Au+Au collisions
at $\sqrt{s_{NN}}$ = 200 GeV. The two distributions are normalized
in the radius region of the TPC gas at $60 < r < 100$ cm. Right:
Conversion probability correction factor for $\pi^0$ as a function
of $p_T$.}
\end{figure*}
Figure \ref{fig:peak_pos_width_200gev} shows the extracted $\pi^0$
peak positions and widths as a function of $p_T$ using different
$\pi^0$ reconstruction methods. Results from real data are compared
to Monte Carlo (MC) simulation of $\pi^0$ embedded in real data. The
$\pi^0$ peak position shows some $p_T$ dependence at low $p_T$ for
the EMC-TPC method and at higher $p_T$ for the EMC-EMC method. At
low $p_T$ the drop is understood as the effect of energy loss of
$e^+/e^-$ tracks due to bremsstrahlung. At high $p_T$ the rise of
peak position as a function of $p_T$ is due to the saturation of
dynamic range for energy measurement from the BSMD in this data set.
The BSMD read-out saturated when the deposited energy exceeds about
6 GeV, leading to more evenly distributed energies of two spatially
close photons when they hit the same BEMC tower, and therefore
produces a larger invariant mass. This effect is more pronounced at
higher $p_T$. The saturation scale was lower than anticipated due to
electronics signal termination issues in 2004. The effect has been
included in the simulation. The trend of $p_T$ dependence is well
reproduced by the simulation but the simulation underestimates the
mass peak position by 4-8\%. The use of TPC photons significantly
improves the $\pi^0$ peak resolution. For the same HT data sample
the EMC-TPC method yields peak widths narrower than those from the
EMC-EMC method, which is consistent with the MC simulation.
Comparing the MB and HT data samples, the requirement of BSMD hits
improves the spatial resolution of EMC photons, and thus measures a
narrower $\pi^0$ peak width.

\subsection{$\mathbf{\pi^0}$ Detection Efficiency} \label{sec:efficiency}

\begin{figure}
\includegraphics[width=3in]{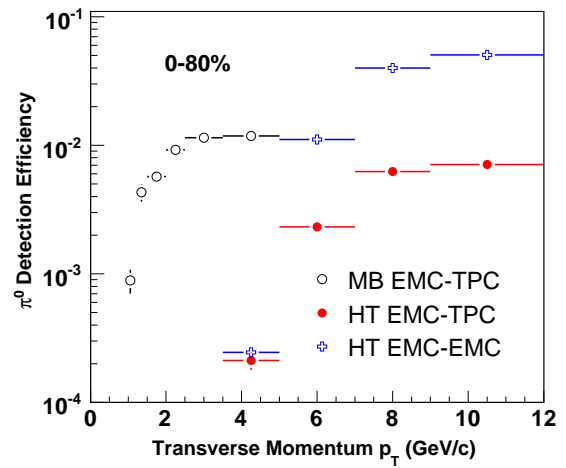}
\caption{\label{fig:pi0_eff_200gev}The overall detection efficiency
of $\pi^0$ from embedding study in 0-80\% Au+Au collisions at
$\sqrt{s_{NN}}$ = 200 GeV.}
\end{figure}
The raw yield of $\pi^0$ is corrected for an overall centrality
dependent detection efficiency calculated from a full MC simulation,
embedding $\pi^0$'s into real events. The efficiency is calculated
as the ratio of the reconstructed $\pi^0$ $p_T$ spectrum over the
input spectrum, using the same cuts as the real data analysis. The
input $\pi^0$ spectrum for the embedding analysis is weighted so
that it reproduces the previously measured charged
\cite{starchargedpion200} or neutral pion spectrum
\cite{phenixpi0200}. The calculated efficiency takes into account
the losses due to acceptance, photon conversion probability,
tracking inefficiency, track energy loss, and track quality cuts.

The conversion probability is crucial for the $\pi^0$ efficiency
calculation when TPC photons are used. A comparison of distributions
for the photon conversion radii between data and MC simulation is
shown in the left panel of Fig. \ref{fig:geo_effect}. The two
distributions are normalized to the data in the inner radius region
of the TPC gas where geometry is simple and well understood. The
comparison indicates that the photon conversion probability can be
well reproduced in the regions of TPC gas and the inner field cage,
but is underestimated in the regions of SVT and SSD where structures
are complex. Similar observation has been reported in earlier
publications \cite{starpi0130}. The results for TPC photons are
corrected for this effect, using the conversion rates in the TPC gas
as a reference. A correction factor
$F_{geo}=(n_{det}/n_{gas})_{data}/( n_{det}/n_{gas})_{MC}$ is
calculated as a function of the conversion photon $p_T$, where
$n_{det}$ and $n_{gas}$ are numbers of conversion points in the
whole detector and in the TPC gas only. In the embedding analysis a
reconstructed TPC photon associated with a MC photon is weighted by
the factor $F_{geo}$ corresponding to its $p_T$, which folds the
correction in the efficiency calculation. The final correction
factor for $\pi^0$ as a function of $p_T$ is shown in the right
panel of Fig. \ref{fig:geo_effect}.

Figure \ref{fig:pi0_eff_200gev} shows the overall detection
efficiency as a function of $p_T$ in Au+Au collisions at
$\sqrt{s_{NN}}$ = 200 GeV. The use of TPC photons is statistically
challenging due to the relatively low conversion probability. Using
EMC photons enhances the efficiency significantly and is preferable
in studying the $\pi^0$ spectrum at high $p_T$. The efficiencies
shown here have taken the conversion probability correction into
account.

\subsection{Systematic Errors} \label{sec:syserror}

Major sources of systematic errors for the $\pi^0$ measurement in
Au+Au collisions at $\sqrt{s_{NN}}$ = 200 GeV are listed in Table
\ref{tab:sys_error_200}. Systematic errors are calculated for each
$p_T$ bin, and systematic errors from different sources are added in
quadrature. The systematic errors are estimated by using several
methods. First, we have varied photon reconstruction cuts and
compared the fully corrected spectra. By changing the geometrical
cuts applied to the TPC photon reconstruction, a systematic error of
10-20\% in the final spectra is obtained. For the EMC photons in MB
events, different energy cuts are used to select EMC photon samples
with different levels of purity. In HT events, instead of a single
tower, a cluster with up to $2\times2$ towers is used to reconstruct
an EMC photon. These various photon reconstruction methods give a
systematic error of 10-20\% to the final $\pi^0$ spectra. Next, We
have also included the uncertainties in the absolute energy scale of
the BEMC, which could affect the overall shape of the $\pi^0$
spectra. An offset in the BEMC energy scale would contribute to the
small deviations in the $\pi^0$ mass peaks between real data and
simulations. We have included a uncertainty of $\pm5\%$
\cite{energyscale} in the BEMC energy scale in our Monte Carlo
simulations. We estimated systematic Errors of 20-35\% throughout
the $p_T$ range. Third, we have varied the fitting procedure used to
extract the raw $\pi^0$ yield. The raw yield of $\pi^0$ depends on
the background fitting function, fit range, and the normalization of
same- and mixed-event invariant mass distributions. Results using
different fitting parameters indicate a systematic error of 10-15\%.
We have also cross checked the uncertainty due to the conversion
probability correction by applying the correction factor as a
function of conversion point position. The result agrees with the
original within 10\%. As a result, a 10\% systematic error is
assigned for the conversion probability correction factor.

\begin{table*}
\caption{\label{tab:sys_error_200}Summary of main sources of
systematic uncertainties on the $\pi^0$ yields in Au+Au collisions
at $\sqrt{s_{NN}}$ = 200 GeV. Systematic uncertainties from varying
photon cuts and yield extraction techniques are $p_T$ uncorrelated,
and systematic uncertainties from the BEMC energy scale and
conversion probability correction are $p_T$ correlated.}
\begin{ruledtabular}
\begin{tabular}{cccc}
 & \multicolumn{1}{c}{MB} & \multicolumn{2}{c}{HT} \\
\cline{2-2}\cline{3-4}
 & EMC-TPC & EMC-TPC & EMC-EMC \\
\hline
Photon Cuts & 10-20\% & 20-30\% & 10-20\% \\
BEMC Energy Scale ($\pm$5\%) & 20-30\% & 25-35\% & 20-35\% \\
Yield Extraction & 10\% & 15\% & 10\% \\
Conversion Probability Correction & 10\% & 10\% & -- \\
\end{tabular}
\end{ruledtabular}
\end{table*}

\section{Results}

\begin{figure}
\includegraphics[width=3.4in]{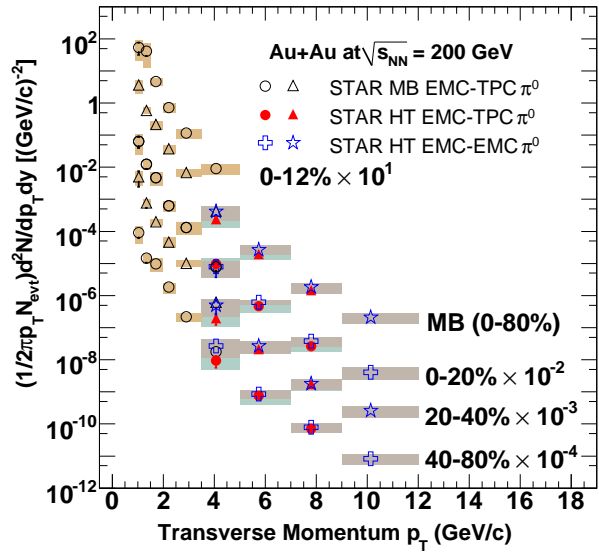}
\caption{\label{fig:spectra_200}Invariant yield of STAR $\pi^0$ as a
function of $p_T$ at mid-rapidity for different collision centrality
bins in Au+Au collisions at $\sqrt{s_{NN}}$ = 200 GeV. Spectra for
different collision centralities are scaled for clarity. Statistical
errors are shown as vertical lines and point-to-point systematic
errors are shown as bars.}
\end{figure}

The $\pi^0$ invariant yield per collision at mid-rapidity
($0<y<0.8$) as a function of $p_T$ in Au+Au collisions at
$\sqrt{s_{NN}}$ = 200 GeV is shown in Fig. \ref{fig:spectra_200}.
Statistical and systematic errors are shown as vertical lines and
bars, respectively. The horizontal size of the vertical bars also
indicates the range of the $p_T$ bin. In addition to the overall MB
0-80\% result, the data sample is also divided into three collision
centrality bins 0-20\%, 20-40\%, and 40-80\% based on measured
charged particle multiplicity at mid-rapidity from the TPC
\cite{centrality}, with 0-20\% the most central collisions. The
$\pi^0$ spectra are measured over an extended $p_T$ range from 1 to
12 GeV/$c$. Results from different $\pi^0$ reconstruction algorithms
and different data samples were compared in overlapping $p_T$
ranges, and were found to be in good agreement. Therefore, in the
following figures only a combined data point using statistical
weighted average of data from various algorithms will be shown in
the overlapping $p_T$ bins.

\begin{figure*}
\includegraphics[width=5in]{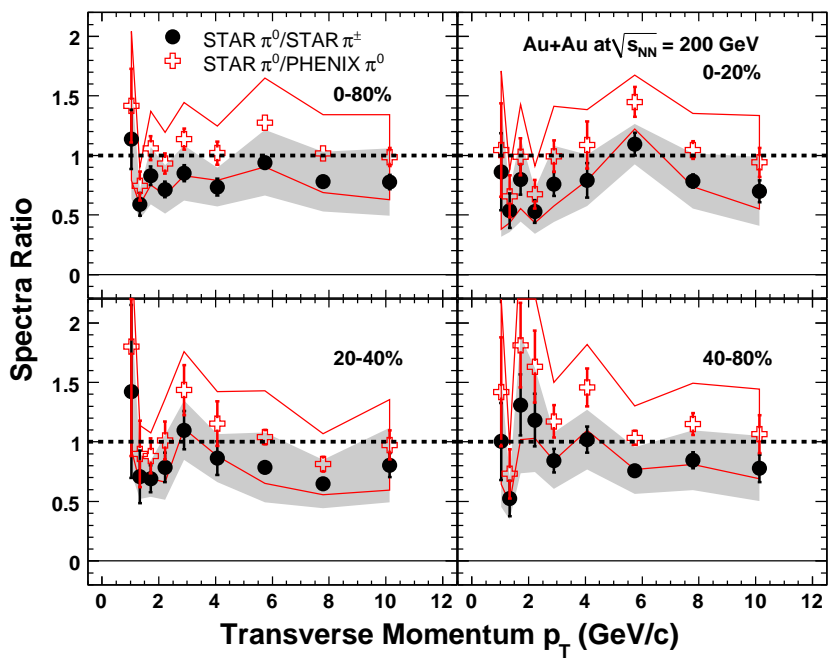}
\caption{\label{fig:ratio_200}The ratios of STAR $\pi^0$ spectra
over $\pi^\pm$ from STAR \cite{starchargedpion200} (solid symbol)
and $\pi^0$ from PHENIX \cite{phenixpi0200} (open symbol) as a
function of $p_T$ for different collision centrality bins in Au+Au
collisions at $\sqrt{s_{NN}}$ = 200 GeV. Statistical errors are
shown as vertical lines and point-to-point systematic errors are
shown as bands.}
\end{figure*}
Our $\pi^0$ spectra are compared to the previously published
$\pi^{\pm}$ and $\pi^0$ results. The ratios of our measured $\pi^0$
spectra to the STAR $\pi^{\pm}$ \cite{starchargedpion200} and the
PHENIX $\pi^0$ \cite{phenixpi0200} in Au+Au collisions at
$\sqrt{s_{NN}}$ = 200 GeV are shown in Fig. \ref{fig:ratio_200}.
Parametrized results of the $\pi^0$ and $\pi^{\pm}$ spectra from
power law functions are used in order to match the $p_T$ binning of
our $\pi^0$ data. The error bars are propagated using the averaged
error of two neighboring data points. The spectrum ratio is slightly
larger in peripheral collisions than in central and mid-central
collisions. With the best statistics in MB 0-80\% collision
centrality, the STAR $\pi^0$ yields are consistent with the PHENIX
$\pi^0$ yields, and about 15\% smaller than the STAR $\pi^{\pm}$
yields over the $p_T$ range. Considering that the systematic
uncertainties in the STAR $\pi^0$ and $\pi^{\pm}$ analyses are
mostly independent, the two yields are consistent within systematic
uncertainties.

The nuclear modification factors can be calculated using peripheral
collisions as a reference ($R_\mathrm {CP}$) or using
nucleon-nucleon collisions as a reference ($R_\mathrm {AA}$):
\[R_{\rm{CP}}(p_T)=\frac{[
d^2N/p_T dy dp_T/ \langle N_{\rm {bin}}\rangle ]^{central}}{[
d^2N/p_T dy dp_T/ \langle N_{\rm {bin}}\rangle ]^{peripheral}},\]
and
\[R_{\rm{AA}}(p_T)=\frac{d^2N_{\rm{AA}}/dy dp_T/\langle N_{\rm
{bin}}\rangle}{d^2\sigma_{pp}/dy dp_T/\sigma_{pp}^{\rm {inel}}},\]
where $\langle N_\mathrm{bin}\rangle$ is the average number of
binary nucleon-nucleon collisions per nucleus-nucleus collision. The
$\sigma_{pp}^{\mathrm {inel}}$ is taken to be 42 mb for
$\sqrt{s_{\mathrm{NN}}}$ = 200 GeV \cite{sigma_pp}. The measurements
of suppression for high $p_T$ charged hadrons from STAR
\cite{starwhitepaper} and neutral pions from PHENIX
\cite{phenixwhitepaper} ($R_\mathrm {CP}$ and $R_\mathrm {AA}$ $<
1$) in most central Au+Au collisions at RHIC provided the first
experimental evidence that partons suffer energy loss in the dense
matter created in these collisions.

\begin{figure}
\includegraphics[width=3in]{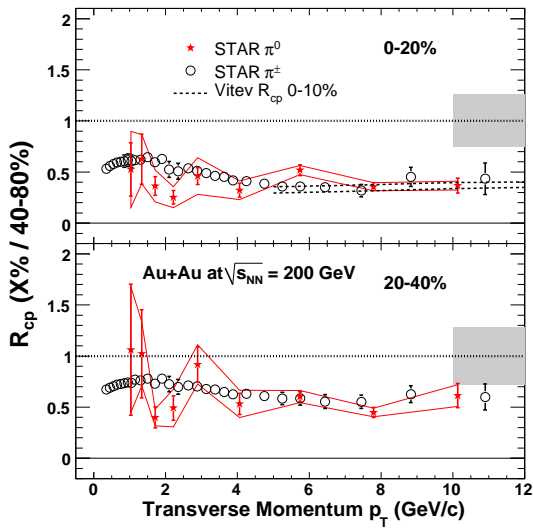}
\caption{\label{fig:rcp_200}The nuclear modification factor
$R_\mathrm{CP}$ as a function of $p_T$ of STAR $\pi^0$ compared to
STAR $\pi^\pm$ \cite{starchargedpion200} in Au+Au collisions at
$\sqrt{s_{NN}}$ = 200 GeV. Statistical errors are shown as vertical
lines and point-to-point systematic errors are shown as solid lines.
The shaded band on the right demonstrates the uncertainty of
$N_\mathrm{bin}$. The dashed curves are jet quenching theoretical
calculations \cite{vitevrcp}.}
\end{figure}
Figure \ref{fig:rcp_200} shows our measurements of the nuclear
modification factor $R_\mathrm{CP}$ for $\pi^0$ as a function of
$p_T$ in Au+Au collisions at $\sqrt{s_{NN}}$ = 200 GeV for the
0-20\% and 20-40\% over 40-80\% collision centrality bins. When
calculating $R_\mathrm{CP}$ some systematic uncertainties cancel
out, such as the BEMC energy scale and conversion probability
correction. Compared to the 40-80\% peripheral Au+Au collisions, the
more central collisions show a suppression of the $\pi^0$ yield
indicated by $R_\mathrm{CP}<1$ and the suppression is even stronger
for the most central collisions. At high $p_T>4$ GeV/$c$ the $\pi^0$
$R_\mathrm{CP}$ is independent of $p_T$ within uncertainties. Our
measured $\pi^0$ $R_\mathrm{CP}$ values show the same magnitude of
suppression as the STAR $\pi^{\pm}$ data \cite{starchargedpion200},
which are shown as open circles.

\begin{figure}
\includegraphics[width=3in]{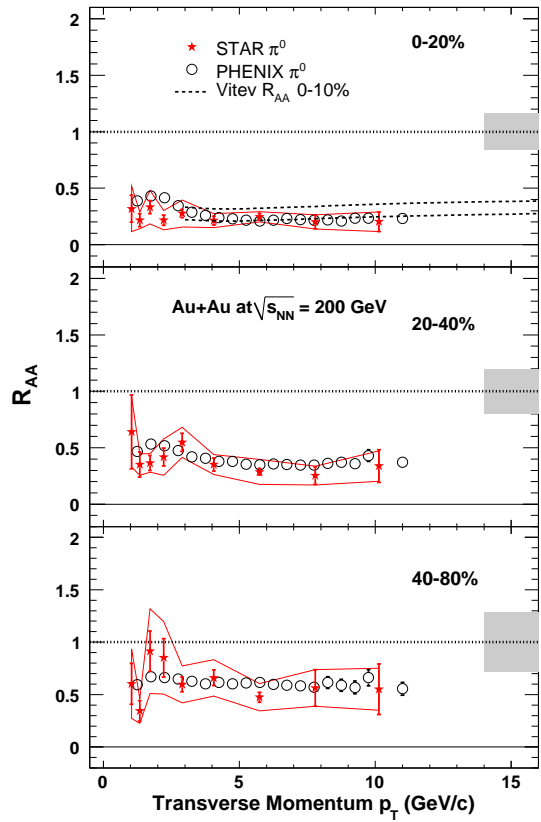}
\caption{\label{fig:raa_200}The nuclear modification factor
$R_\mathrm{AA}$ as a function of $p_T$ of STAR $\pi^0$ compared to
PHENIX $\pi^0$ \cite{phenixpi0200} in Au+Au collisions at
$\sqrt{s_{NN}}$ = 200 GeV. Statistical errors are shown as vertical
lines and point-to-point systematic errors are shown as solid lines.
The shaded band on the right demonstrates the uncertainties of
$N_\mathrm{bin}$ and the normalization uncertainty in $p+p$
collisions of 14\% \cite{starppnorm}. The dashed curves are
theoretical calculations in 0-10\% Au+Au collisions
\cite{vitevraa}.}
\end{figure}
Figure \ref{fig:raa_200} shows our measurements of the nuclear
modification factor $R_\mathrm{AA}$ for $\pi^0$ as a function of
$p_T$ in Au+Au collisions at $\sqrt{s_{NN}}$ = 200 GeV for 0-20\%,
20-40\%, and 40-80\% collision centrality bins, where a
parameterized description of the $\pi^{\pm}$ spectrum in $p+p$
collisions at $\sqrt{s_{NN}}$ = 200 GeV from Ref.
\cite{star_chargedpi_pp} is used to calculate $R_\mathrm{AA}$. The
error bars of the $p+p$ result are propagated using the averaged
error of two neighboring data points. The $\pi^0$ $R_\mathrm{AA}$
shows a similar centrality dependence as the $R_\mathrm{CP}$. In the
most central Au+Au collisions the $\pi^0$ yield is suppressed by a
factor of about 5 relative to the expectation from scaled
nucleon-nucleon collisions. For all the collision centrality bins,
our measured $R_\mathrm{AA}$ values for $\pi^0$ at high $p_T$ agree
with previously published results from the PHENIX collaboration
\cite{phenixpi0200} within systematic uncertainties.

The nuclear modification factors for inclusive light quark mesons at
high $p_T$ in central heavy-ion collisions have been investigated
with several model calculations. The nuclear modification factors
$R_\mathrm{CP}$ and $R_\mathrm{AA}$ have been calculated in terms of
parameters such as the initial gluon density \cite{vitevrcp,
vitevraa} and the medium transport coefficient $\hat{q}$ \cite{PQM},
which characterize properties of the dense matter created. Recent
theoretical calculations suggest that the collisional energy loss
may also play an important role in explaining the large suppression
of non-photonic electrons from heavy quark decays \cite{star_npe}.
In Fig. \ref{fig:rcp_200}, we show an example of a theoretical
calculation of $R_\mathrm{CP}$ with initial gluon density $dN^g/dy$
= 1150 in 0-10\% Au+Au and between 100 and 150 in 40-80\% Au+Au
collisions \cite{vitevrcp}. In Fig. \ref{fig:raa_200}, theoretical
calculations with $dN^g/dy$=800 to 1150 for 0-10\% Au+Au collision
centrality \cite{vitevraa} are shown as dashed curves in comparison
to measurements from STAR and PHENIX from 0-20\% collision
centrality. Experimental measurements and theoretical predictions
agree reasonably well, indicating that the yield suppression of
light quark mesons may be accounted for by the parton energy loss
mostly through gluon radiation.

\section{Summary}

We have presented the first STAR results for $\pi^0$ production in
Au+Au collisions at $\sqrt{s_\mathrm{NN}}$ = 200 GeV. The $\pi^0$
spectra are measured over the range of 1$<p_T<$12 GeV/$c$ using the
combination of conversion photons from TPC reconstruction and
photons from BEMC energy measurement. Despite the relatively large
tower size, the STAR BEMC alone can be used to reconstruct $\pi^0$'s
for $p_T>4$ GeV/$c$. The use of conversion photons significantly
enhances detection capability for $\pi^0$'s at low and intermediate
$p_T$ and extends our $\pi^0$ measurement to a lower $p_T$ range.

Our measurements of $\pi^0$ spectra are consistent with the STAR
charged $\pi^{\pm}$ and PHENIX $\pi^0$ results within statistical
and systematic errors. The nuclear modification factors
$R_\mathrm{CP}$ and $R_\mathrm{AA}$ of the STAR $\pi^0$ data confirm
the previously published $\pi$ results and can be described by
theoretical calculations based on parton energy loss through gluon
radiation in the dense medium created at RHIC. In the most central
Au+Au collisions the inclusive $\pi^0$ yield shows a factor of about
5 suppression relative to the expectation from scaled $p+p$
collisions for $p_T>5$ GeV/$c$. Our measurements confirm the
magnitude of light hadron suppression observed in central Au+Au
collisions and provide further support for the physical picture of
jet quenching in the dense matter created in nucleus-nucleus
collisions at RHIC.

\section*{Acknowledgments}

We thank the RHIC Operations Group and RCF at BNL, and the NERSC
Center at LBNL and the resources provided by the Open Science Grid
consortium for their support. This work was supported in part by the
Offices of NP and HEP within the U.S. DOE Office of Science, the
U.S. NSF, the Sloan Foundation, the DFG cluster of excellence
`Origin and Structure of the Universe', CNRS/IN2P3, RA, RPL, and EMN
of France, STFC and EPSRC of the United Kingdom, FAPESP of Brazil,
the Russian Ministry of Sci. and Tech., the NNSFC, CAS, MoST, and
MoE of China, IRP and GA of the Czech Republic, FOM of the
Netherlands, DAE, DST, and CSIR of the Government of India, the
Polish State Committee for Scientific Research,  and the Korea Sci.
\& Eng. Foundation


\begin{thebibliography}{0}

\bibitem{starwhitepaper} J. Adams {\it et al.}, Nucl. Phys. {\bf A757}, 102 (2005).

\bibitem{phenixwhitepaper} K. Adcox {\it et al.}, Nucl. Phys. {\bf A757}, 184 (2005).

\bibitem{starppnorm} J. Adams {\it et al.}, Phys. Rev. Lett. {\bf 91}, 172302
(2003).

\bibitem{highptsuppression} S. S. Adler {\it et al.}, Phys. Rev. Lett. {\bf 91}, 072301
(2003); B. B. Back {\it et al.}, Phys. Lett. B {\bf 578}, 297
(2004); I. Arsene {\it et al.}, Phys. Rev. Lett. {\bf 91}, 072305
(2003).

\bibitem{noawayside} C. Adler {\it et al.}, Phys. Rev. Lett. {\bf 90}, 082302
(2003); J. Adams {\it et al.}, Phys. Rev. Lett. {\bf 97}, 162301
(2006).

\bibitem{GLV} M. Gyulassy, P. Levai, and I. Vitev, Phys. Rev. Lett. {\bf 85}, 5535
(2000); M. Gyulassy {\it et al.}, in: R. C. Hwa, X.-N. Wang (Eds.),
Quark Gluon Plasma 3, World Scientific, Singapore, 2003, p. 123,
nucl-th/0302077.

\bibitem{directphoton} S. S. Adler {\it et al.}, Phys. Rev. Lett. {\bf 94}, 232301 (2005).

\bibitem{bemc} M. Beddo {\it et al.}, Nucl. Instrum. Methods A {\bf 499}, 725 (2003).

\bibitem{tpc} M. Anderson {\it et al.}, Nucl. Instrum. Methods A {\bf 499}, 659 (2003).

\bibitem{starpi0130} J. Adams {\it et al.}, Phys. Rev. C {\bf 70}, 044902 (2004).

\bibitem{startrigger} F. S. Bieser {\it et al.}, Nucl. Instrum. Methods
A {\bf 499}, 766 (2003).

\bibitem{starchargedpion200} B. I. Abelev {\it et al.}, Phys. Rev. Lett. {\bf 97}, 152301
(2006).

\bibitem{phenixpi0200} A. Adare {\it et al.}, Phys. Rev. Lett. {\bf 101}, 232301 (2008).

\bibitem{energyscale} B. I. Abelev {\it et al.}, Phys. Rev. Lett. {\bf 97}, 252001 (2006).

\bibitem{centrality} J. Adams {\it et al.}, Phys. Rev. Lett. {\bf 92}, 112301 (2004).

\bibitem{sigma_pp} C. Amsler {\it et al.}, Phys. Lett. B {\bf 667}, 364 (2008).

\bibitem{star_chargedpi_pp} J. Adams {\it et al.}, Phys. Lett. B {\bf 637}, 161
(2006); J. Adams {\it et al.}, Phys. Lett. B {\bf 616}, 8 (2005).

\bibitem{vitevrcp} I. Vitev, Phys. Lett. B {\bf 639}, 38 (2006).

\bibitem{vitevraa} I. Vitev and M. Gyulassy, Phys. Rev. Lett. {\bf 89}, 252301 (2002).

\bibitem{PQM} A. Dainese, C. Loizides, and G. Paic, Eur. Phys. J. C
{\bf 38}, 461 (2005); C. Loizides, Eur. Phys. J. C {\bf 49}, 339
(2007).

\bibitem{star_npe} B. I. Abelev {\it et al.}, Phys. Rev. Lett. {\bf 98}, 192301
(2007).

\end{thebibliography}

\end{document}